\documentstyle[preprint,aps]{revtex}
\begin{document}
\newcommand\beq{\begin{equation}}
\newcommand\eeq{\end{equation}}
\newcommand\bea{\begin{eqnarray}}
\newcommand\eea{\end{eqnarray}}
\renewcommand\aa{{\cal A}}
\tightenlines
\draft
\title {Effective Potential for the Second Virial Coefficient at
  Feshbach Resonance}
\author{Rajat K. Bhaduri$^1$ and M. K. Srivastava$^2$}
\address{1. Department of Physics and Astronomy, McMaster University,
Hamilton, Canada L8S 4M1}
\address{2. Institute Instrumentation Centre, I.I.T., Roorkee 247667, India}
\maketitle

\begin{abstract}

In a recent paper entitled ``High temperature expansion applied to
fermions near Feshbach resonance'', (Phys. Rev. Lett. {\bf 92} 
160404 (2004)), Ho and Mueller have demonstrated a remarkable similarity
between its high and low temerature properties at resonance. 
%Specifically, they
%examine the interaction energy of a high-temperature Fermi gas in the 
%Boltzmann regime in terms
%of the quantum second virial coefficient, and compare it with 
%the experimental data at low temperatures. 
The quantum second virial coefficient plays a crucial role in
their analysis, and has a universal value at resonance.
In this paper, we explore the connection between the
quantum and classical second virial coefficients, and show that near a  
Feshbach resonance an exact mapping from the quantum to classical form
is possible. This gives rise to a scale-independent inverse square
effective potential for the classical virial coefficient. It is
suggested that this may be tested by measuring the isothermal 
compressibility of the gas on the repulsive side of the resonance..

\end{abstract}
\vskip .5 true cm
\pacs{PACS:~03,75.-b,~03.75.Ss}

\maketitle
\narrowtext

Great advances have been made in the study of ultracold trapped atoms
in recent times. The interaction between the atoms is normally weak
in a dilute gas, but may be enhanced drastically by making use of the
so-called Feshbach resonance~\cite{feshbach}. 
This may be achieved by applying a
magnetic field to tune the energy of the Zeeman level of the trapped
atoms to one of the molecular resonances. At a zero-energy resonance,
the scattering phase shift is $\pi/2$, and the cross section reaches the
unitary limit. It is argued that the properties of a gas at Feshbach
resonance have a universal character~\cite{uni}, irrespective of the 
type of gas
in the trap. In this connection, Ho and Mueller~\cite{ho} recently 
considered the interaction
energy of a two-component Fermi gas near a Feshbach resonance. 
They compared the available experimental data of the    
ultracold gas in the vicinity of a Feshbach resonance with the
predictions in the Boltzmann regime using the {\it quantum} second virial 
coefficient. Surprisingly, it was found that the high temperature
expression, when used at micro-Kelvin temperatures, explains the data 
reasonably well. In the present work, we examine the use of the quantal
second virial coefficient in more detail, and its connection to its
classical counterpart at the Feshbach resonance. We find that the
quantal expression may be mapped onto the classical formula exactly in
the vicinity of the Feshbach resonance using an inverse square
interaction in the configuration space.          

To appreciate this in more detail, note that the virial expansion of 
a (one-component) gas (classical as well as quantal) for the pressure $P$ at a
temperature $T$ is given by~\cite{pathria} 
\beq 
\frac{P}{n\tau}=1 + a_2(n\lambda^3) + a_3(n\lambda^3)^2 +....~, 
\eeq
where $a_2, a_3$... etc are the dimensionless second, third..virial 
coefficients, $\tau=k_BT$,  $n=N/V$ is the number density of 
particles, and 
$\lambda=\sqrt{2\pi\hbar^2/m\tau}$ is the thermal de Broglie wavelength.
The free energy $F=E-TS$ may be easily obtained by integrating the
above $P$ with respect to the volume $V$ (since
$P=-(\partial F/\partial V)_{\tau}$), and hence also the energy 
$E=-\tau^2[\partial (F/\tau)/\partial\tau]$. After
subtracting out the energy of the perfect gas part
(classical or quantal), one obtains the virial series for the 
interaction energy :  
\beq
\frac{E_{int}}{V}=\frac{3}{2}n\tau\left[(n\lambda^3)\{a_2-\frac{2}{3}
\tau\frac{da_2}{d\tau}\}+(n\lambda^3)^2\{a_3-\frac{2}{3}
\tau\frac{da_3}{d\tau}\frac{1}{2}\}+...
(n\lambda^3)^{(j-1)}\{a_j-\frac{2}{3}
\tau\frac{da_j}{d\tau}\frac{1}{j-1}\}+...\right]
\label{interaction}
\eeq
This expression is valid for a classical as well as a quantum gas,
provided the appropriate classical or quantal virial coefficients are
used. Whereas for a classical gas the virial coefficients may be
expressed as integrals involving the interaction potential, for the
quantum problem a solution of the j-body problem is needed to obtain
$a_j$. Henceforth we shall denote the classical virial coefficients by 
$\aa_j$ to differentiate from their quantum counterparts,  
to be still denoted as $a_j$.  
For an interesting example where classical and quantum results
are very different, consider a hard sphere classical Boltzmann gas.   
For a hard-sphere diameter $r_c$, 
${\cal A}_2=\frac{2\pi}{3}(\frac{r_c}{\lambda})^3$, and the higher order 
${\cal A}_j$'s may be expressed as ${\cal A}_j=\kappa ({\cal
  A}_2)^{j-1}$, where $\kappa$ is a constant. Substituting this 
in Eq.(\ref{interaction}), we see that the interaction energy of the
classical hard sphere gas vanishes identically in each power of 
$(n\lambda^3)$. This, of course, is not the case for a bosonic or
fermionic quantum gas at low temperatures.   

We note that for a classical gas, $(n\lambda^3)<<1$, and the series 
(\ref{interaction}) may be terminated after the first order term  
in most cases:
\beq
\frac{E_{int}}{V}=\frac{3}{2}n\tau\left[(n\lambda^3)\{a_2-\frac{2}{3}
\tau\frac{da_2}{d\tau}\}\right]~.
\label{term}
\eeq
The classical second virial coefficient $\aa_2$ is given by~\cite{pathria} 
\beq
\aa_2(\beta)=\frac{2\pi}{\lambda^3}\int_0^{\infty}dr~ r^2 
\left (1-\exp(-\beta V(r)\right)~,
\label{class}
\eeq
where $\beta=1/\tau$, and $V(r)$ is the interatomic two-body potential.
The quantum second virial
coefficient requires a knowledge of the bound-state and the continuum 
spectra. For a one-component quantum gas, $b_2=-a_2$ is given by~\cite{pathria} 
\bea
b_2(\beta)&=&\sum_{l=0}^{\infty}(2l+1)b_2^{(l)}\\
b_2^{(l)}(\beta)&=&2 \sqrt{2}\left[\sum_n \exp(-E_{n,l}\beta)+\frac{1}{\pi}
\int_0^{\infty}dk~\frac{\partial\delta_l}{\partial k} 
\exp(-\frac{\hbar^2 k^2\beta}{m})\right]~.
\label{delta}
\eea 
In the above, $E_{n,l}$'s are the two-body bound states (if any), and
$\delta_l$ is the scattering phase shift in the $l$th partial wave. 
In the quantum case, for a Fermi gas of atoms interacting with a delta-function
pseudo-potential, the spin-up atoms interact only with the
spin-down ones, and one may consider a two-component (up, down) gas
with $n_{\uparrow}=n_{\downarrow}=n/2$. In this situation, $n^2$ in
(\ref{term}) should be replaced by $n^2/4$. Often, the
multiplicative prefactor in Eq.(\ref{delta}) is omitted, being
transferred instead to the RHS of Eq.(\ref{term}). The modified
equation for a two-component quantum gas then reads 
\beq
\frac{E_{int}}{V}=\frac{3}{2}n\tau\left[(n\lambda^3)\{-\frac{b_2}{\sqrt{2}}
+\frac{\sqrt{2}}{3}\tau\frac{db_2}{d\tau}\}\right]~,
\label{term2}
\eeq
with
\beq
b_2^{(l)}(\beta)=\left[\sum_n \exp(-E_{n,l}\beta)+\frac{1}{\pi}
\int_0^{\infty}dk~\frac{\partial\delta_l}{\partial k} 
\exp(-\frac{\hbar^2 k^2\beta}{m})\right]~.
\label{delta2}
\eeq
Ho and Mueller~\cite{ho} used the above formula for the quantum virial coefficient in the 
high temperature expression (\ref{term2}) to 
explain the low-temperature data. Two questions immediately arise
regarding this treatment :\\
(a) For the quantum gas, since $(n\lambda^3)$ 
is larger than, or of order unity, higher order terms in
Eq.(\ref{interaction}) may be important. Presumably, near the Feshbach 
resonance, it is still the two-body term that dominates the
interaction energy.\\     
(b) One may also argue that while using the high-temperature
expression (\ref{term2}) for the interaction energy, one should take
the classical Eq.(\ref{class}) for the second virial coefficient
coefficient, instead of Eq.(\ref{delta2}). 

In the following, we attempt to answer the second question in some
detail. Note that at low temperatures (large $\beta$), only small
values of $k$ contribute dominantly in the integrand in
Eq. (\ref{delta}), and therefore only the $l=0$ partial wave in the
sum is important. Therefore, rather than the total classical $\aa_2$
given by Eq. (\ref{class}), we should consider its $l=0$ classical
component. Making the decomposition 
\beq
\aa_2(\beta)=\sum_{l=0}^{\infty}(2l+1) \aa_2^{(l)}~,
\eeq
it may be verified easily (by replacing the sum over $l$ by an
integral), that 
\beq
\aa_2^{(l)}=\frac{1}{\lambda}\int_0^{\infty}~ dr~ 
\exp\left[-\frac{\hbar^2l(l+1)}{mr^2}\beta\right][1-\exp(-\beta V(r))]~.
\label{partclass}
\eeq
One may enquire about the relation between the ``semiclassical''
expression (\ref{partclass}) and the quantam counterpart
$-b_2^{(l)}$. This question was actually answered nearly seventy years
back by Kahn~\cite{kahn} in his doctoral dissertation. We shall give here a new
compact derivation of his result, and for simplicity consider a
potential with no bound states present. Using
(\ref{delta2}), and the relation $E=\frac{\hbar^2 k^2}{m}$, we have    
\beq  
b_2^{(l)}(\beta)= \left[\frac{1}{\pi}
\int_0^{\infty}dk~\frac{\partial\delta_l}{\partial k} 
\exp(-\frac{\hbar^2 k^2\beta}{m})\right]~=
\left[\frac{1}{\pi}
\int_0^{\infty}dE~\frac{\partial\delta_l}{\partial E} 
\exp(-\beta E)\right]~.
\label{ho}
\eeq
With this definition of the virial coefficient, at a zero energy 
resonance, $b_2^{(0)}=1/2$ or $-1/2$, according as the scattering
length is negative (attractive interaction), or positive (repulsive 
interaction). We also note that $b_2^{(l)}(\beta)$ is just the
Laplace transform of the derivative of the phase shift with rspect to
the energy. In the lowest order WKB approximation, the phase shift is
given by~\cite{messiah}
\beq
[\delta_l(E)]_{WKB}=\sqrt{\frac{m}{\hbar^2}}\left[\left(\int_{r_l}^{\infty}
\sqrt{E-V_l(r)}-\int_{r_0}^{\infty}
\sqrt{E-\frac{\hbar^2 l(l+1)}{mr^2}}~\right) dr\right]~.
\label{wkb}
\eeq
In the above, the effective potential $V_l(r)$ is defined as 
\beq
V_l(r)=V(r)+\frac{\hbar^2l(l+1)}{mr^2}~.
\eeq 
In Eq.(\ref{wkb}), $r_l$ and $r_0$ are the classical turning
points where the respective integrands go to zero. The derivative of
the phase shift may be written as 
\beq
[\frac{d\delta_l}{dE}]_{WKB}=\frac{1}{2}\sqrt{\frac{m}{\hbar^2}}\left[\left(
\int_0^{\infty}\frac{\Theta(E-V_l(r))}{\sqrt{E-V_l(r)}}-
\int_0^{\infty}\frac{\Theta(E-{\hbar^2l(l+1)\over {2mr^2}})}
{\sqrt{E-{\hbar^2l(l+1)\over {mr^2}}}}\right)dr\right]~
\label{wkb1}
\eeq
in terms of the unit step function $\Theta(x)$. By noting that the
Laplace transform of $\frac{\Theta(E-\mu)}{\sqrt{E-\mu}}$ is 
$\sqrt{\frac{\pi}{\beta}}\exp(-\beta \mu)$, we immediately obtain the
desired result
\beq
[b_2]_{WKB}^{(l)}=-\frac{1}{\sqrt{2}\lambda}\int_0^{\infty}~dr
\exp\left[-\beta\frac{\hbar^2l(l+1)}{mr^2}\right]
[1-\exp(-\beta V(r))]~~=~-\frac{\aa_2^{(l)}}{\sqrt{2}}~.
\label{nogo}
\eeq
The multiplicative factor of $\frac{1}{\sqrt{2}}$ may be understood by 
noting that $b_2^{(l)}$ as defined by Eq.(\ref{delta2}), is the quantum
canonical two-body partition function for the {\it relative} motion in
the $l^{th}$ partial wave, {\it without} the noninteracting continuum part.  
Noting that the radial motion in a given partial wave is one 
dimensional, the classical limit of $b_2^{(l)}$ is  
\bea
Z_{cl}^{(l)}&=&\frac{1}{h}\int_{-\infty}^{\infty}dp_r
\exp(-\frac{p_r^2\beta}{m})
\int_0^{\infty}dr \{\exp[-\beta (V(r)+\frac{\hbar^2 l(l+1)}{mr^2})]-
            \exp[-\beta (\frac{\hbar^2 l(l+1)}{mr^2})]\}~,\\
&=&-\frac{1}{\sqrt{2}\lambda}\int_0^{\infty}~dr
\exp\left[-\beta\frac{\hbar^2l(l+1)}{mr^2}\right]
[1-\exp(-\beta V(r))]~~=~-\frac{\aa_2^{(l)}}{\sqrt{2}}~.
\label{kol}
\eea

If the lowest order WKB approximation (\ref{wkb1}) for a given potential  
 were exact, the classical and quantum (partial wave)
virial coefficients would be related by
$b_2^{(l)}=-\aa_2^{(l)}/\sqrt{2}$. 
But in general, specially near
the Feshbach resonance, the WKB approximation for the phase shift is 
not accurate. Moreover, the universality property of the quantum 
$b_2^{(0)}=\pm 1/2$
at the resonance is not reflected by its classical or WKB counterpart. 
The latter invariably depends on temperature as well as the parameters
of the potential (even at resonance). 
As an illustration, consider an attractive square-well potential of depth
$-V_0$, and range $R$ for the two-body interaction. When the
dimensionless ``strength factor'' $s=\frac{4}{\pi^2}
\frac{mV_0}{\hbar^2}R^2$ is unity, there is an s-wave zero-energy bound state,
with $b_2^{(0)}=\frac{1}{2}$. If, on the other hand, we were to use the
WKB expression (\ref{nogo}), we would obtain
$[b_2^{(0)}]_{WKB}=\frac{R}{\sqrt{2}\lambda}(\exp(\beta V_0)-1)$. Unlike the
exact quantum result, this depends both on temperature and the 
potential parameters, so that universality is lost.

The only exception to this rule
is for a scale-independent inverse square potential of the type 
\beq
V(r)=\frac{\hbar^2}{m}\frac{\alpha^2}{r^2}~,
\label{calogero}
\eeq
where $\alpha$ is a dimensionless (real) parameter. It is known
that the WKB method gives the exact phase shift~\cite{exact} for such a potential,
and it is easy to check that the classical virial coefficient,  
evaluated using Eq.(\ref{partclass}), is 
\beq
\left(\aa_2^{(0)}\right)_{res}=\frac{|\alpha|}{\sqrt{2}}~.
\label{kyabat} 
\eeq
Using Eq.(\ref{kol}) we see that for $\alpha^2=1$, 
we get $b_2^{(0)}=-1/2$, as is appropriate for a repulsive potential at 
a zero-energy resonance.

We now show that the
classical form of the second virial coefficient may also be recovered
from the quantum expression through an exact mapping near the
zero-energy resonance. To this end, we start with Eq.(\ref{ho}) for
the $l=0$ partial wave, 
\beq  
b_2^{(0)}(\beta)= \left[\frac{1}{\pi}
\int_0^{\infty}dk~\frac{d\delta_0}{dk} 
\exp(-\frac{\hbar^2 k^2\beta}{m})\right]~.
\label{map}
\eeq
Near zero-energy ($k\rightarrow0$), we may use the shape-independent effective
range expansion, and for $l=0$ it is given by 
\beq
k \cot\delta_0=-\frac{1}{a}+\frac{1}{2} r_0^2 +...
\eeq
In the above, $a$ is the scattering length that approaches
$\pm\infty$ near a resonance, and $r_0$ is the effective range. The
virial coefficient near the resonance is governed by $a$ alone, and we
may ignore the second term. In that case, using (\ref{map}), we get 
\beq
b_2^{0}=-\frac{1}{\pi}\int_0^{\infty}~dk \frac{a}{1+a^2k^2}~
\exp\left(-\frac{\hbar^2k^2\beta}{m}\right)~.
\eeq
This may be rewritten as 
\beq
b_2^{0}=\frac{1}{\pi}\int_0^{\infty}~dk \frac{a}{1+a^2k^2}~\left(1-
\exp\left(-\frac{\hbar^2k^2\beta}{m}\right)~\right)~-
\frac{1}{\pi}\int_0^{\infty}~dk \frac{a}{1+a^2k^2}~.
\eeq
The second term yields $-sgn(a)~\frac{1}{2}$. As $a\rightarrow\pm\infty$, we
may then write  
\beq
b_2^{(0)}\simeq \frac{sgn(a)}{|a|\pi}\int_0^{\infty} \frac{dk}{k^2}
\left(1-
\exp\left(-\frac{\hbar^2k^2\beta}{m}\right)~\right)~-sgn(a)~\frac{1}{2}
\eeq
By making the transformation $k=\alpha/x$, we get 
\beq
b_2^{(0)}\simeq \frac{sgn(a)}{|a|\pi\alpha}\int_0^{\infty}~
\left(1-
\exp\left(-\frac{\hbar^2\alpha^2\beta}{m x^2}\right)~
\right)~dx~-~sgn(a)~\frac{1}{2}
\eeq
We have already noted, from Eq.(\ref{kyabat}), that the second term on
the RHS may be
reproduced classically by an inverse square potential with 
$\alpha=1$ for the repulsive case. The first term on the RHS, on 
performing the integral, is independent of $\alpha$, and gives a
correction of order $\hbar$ to the classical value in the
neighbourhood of the resonance. We may thus write the above equation
in the form 
\beq
b_2^{(0)}\simeq sgn(a)~\left(-\frac{\aa_2^{(0)}}{\sqrt{2}}\right)_{res}\left[1-
\frac{\sqrt{2}\lambda}{|a|\pi}\right]~,
\eeq
where $\aa_2^{(0)}$ is obtained from Eq.(\ref{partclass}) with the
inverse square potential (\ref{calogero}) of strength $\alpha=1$. 

We have shown above that as the Feshbach resonance is approached from
the repulsive side, the two-body physics may be simulated by a
universal inverse square potential of unit strength. This does not
appear to hold on the attractive side. The inverse square potential in
one space dimension is known as the Calogero-Sutherland 
potential~\cite{calo}, and
has been thoroughly studied, including its two-point correlation 
function~\cite{krivnov}.  
The integral of the latter is related to the isothermal
compressibility, and can be computed numerically~\cite{raj}. Its experimental
measurement near the Feshbach resonance may be of interest, specially
in view of the findings in this paper. The other intriguing point is
the so-called Efimov effect~\cite{efimov}. When the two-body scattering length
approaches infinity from the {\it attractive side}, a very large
number of three-body bound states appear. More over, Efimov has shown
that the universal effective three-body interaction in this case is
$-\frac{s_0^2\hbar^2}{2mR^2}$, where
$R^2=2(r_{12}^2+r_{23}^2+r_{31}^2)/3$, and $s_0^2\simeq 1$. We find,
from the two-body analysis, that an inverse square two-body effective potential
also arises at Feshbach resonance, albeit on the
repulsive side. More research in this direction is planned.   

We thank Don Sprung for going through the manuscript carefully.
One of the authors (R.K.B.) acknowledges NSERC (Canada) for financial support.

\end{document}